\documentstyle[aps,prl,preprint,tighten]{revtex}

\newcommand{\half}{{\textstyle{1\over2}}}
\newcommand{\halfhalf}{(\half,\half)}
\newcommand{\D}{{\mathcal D}}
\newcommand{\tfrac}[2]{{\textstyle{\frac{#1}{#2}}}}
\newcommand{\Kappa}{\kappa}
\renewcommand{\d}{{\rm d}}
\newcommand{\Tr}{\mbox{Tr\,}}
\newcommand{\Det}{\mbox{{\rm Det}}}
\newcommand{\beq}{\begin{equation}}
\newcommand{\eeq}[1]{\label{#1}\end{equation}}
\newcommand{\bea}{\begin{eqnarray}}
\newcommand{\eea}[1]{\label{#1}\end{eqnarray}}

\newcommand{\rtitle}[1]{\ifpreprintsty{\sl #1},\fi}

\begin{document}
\preprint{\vtop{\hbox{MCTP-01-18}\hbox{hep-th/0105008}\vskip24pt}}

\title{Quantum $M^{2} \rightarrow 2\Lambda/3$ discontinuity for massive gravity 
with a $\Lambda$ term}

\author{M.~J.~Duff, James T.~Liu and H.~Sati\footnote{Email addresses:
\{mduff, jimliu, hsati\}@umich.edu}}

\address{Michigan Center for Theoretical Physics\\
Randall Laboratory, Department of Physics, University of Michigan\\
Ann Arbor, MI 48109--1120, USA}

\maketitle

\begin{abstract}
In a previous paper we showed that the absence of the van 
Dam-Veltman-Zakharov discontinuity as $M^{2}\rightarrow 0$ for massive 
spin-2 with a $\Lambda$ term is an artifact of the tree approximation, 
and that the discontinuity reappears at one loop, as a result of going 
from five degrees of freedom to two.  In this paper we show that a similar 
classical continuity but quantum discontinuity arises in the 
``partially massless'' limit $M^{2}\rightarrow 2\Lambda/3$, as a 
result of going from five degrees of freedom to four.

\end{abstract}


\ifpreprintsty\else   
\begin{multicols}{2}
\fi
\narrowtext

In a previous paper \cite{dilkes}, we showed that the absence 
\cite{Higuchi,Kogan,Porrati} of the van Dam-Veltman-Zakharov 
discontinuity \cite{vdv,zak} for massive spin-2 with a $\Lambda$ term 
is an artifact of the tree approximation, and that the discontinuity 
reappears at one loop.  This result may be understood as follows.  While
a generic massive graviton propagates five degrees of freedom, gauge 
invariance ensures only propagation of the familiar two degrees of 
freedom of a massless graviton.  Although the introduction of 
$\Lambda\ne0$ allows for a smooth classical $M^2\to0$ limit, the 
mismatch between two and five degrees of freedom cannot be eliminated 
altogether, and the discontinuity shows up at the quantum level.

Curiously, the presence of a cosmological constant allows for new
gauge invariances of massive higher spin theories, yielding a rich
structure of ``partially massless'' theories with reduced degrees of
freedom \cite{Deser:2001pe}.  In particular, for spin-2 a single
gauge invariance shows up at the value $M^2=2\Lambda/3$, yielding a
partially massless theory with four degrees of freedom
\cite{deser4,deser,Deser:2001wx}.  In this paper we extend the result
of \cite{dilkes} to the partially massless theory and show that a
discontinuity first arises at the quantum level as $M^2\to2\Lambda/3$.


We work in four dimensions with Euclidean signature \mbox{$(+ + + +)$}.
As in Refs.~\cite{Porrati,dilkes}, we take the massive spin-2 theory to be
given by linearized gravity with the addition of a Pauli-Fierz mass term.
Thus our starting point is the action
\begin{equation}
S[h_{\mu\nu},T_{\mu\nu}] = S_L[h_{\mu\nu}] + S_M[h_{\mu\nu}]
+ S_T[h\cdot T]\, ,
\end{equation}
where $S_L$ is the Einstein-Hilbert action with cosmological constant
$S_E = -\frac{1}{16\pi G} \int\d^4 x \sqrt{\hat g} (\hat R - 2 \Lambda)$,
linearized about a background metric $g_{\mu\nu}$ satisfying the Einstein
condition, $R_{\mu\nu}=\Lambda g_{\mu\nu}$.  Taking $\hat g_{\mu\nu} =
g_{\mu\nu} + \Kappa h_{\mu\nu}$ where $\kappa^2 = 32 \pi G$, this linearized
action for $h_{\mu\nu}$ is
\begin{eqnarray}
\label{SL}
S_L = \int \d^4 x \sqrt{g}
\Bigl[&&\tfrac{1}{2}\tilde h^{\mu\nu}
\left(-g_{\mu\rho}g_{\nu\sigma}\Box-2R_{\mu\rho\nu\sigma}\right)h^{\rho\sigma}
\nonumber\\
&& -\nabla^\rho \tilde h_{\rho\mu} \nabla^{\sigma} \tilde h_\sigma{}^\mu
\Bigr] \, ,
\end{eqnarray}
where $\tilde h_{\mu\nu} = h_{\mu\nu} - \frac{1}{2} g_{\mu\nu}
h^{\sigma}{}_\sigma$.  All indices are raised and lowered with respect
to the metric $g_{\mu\nu}$, and $\nabla_\mu$ is taken to be covariant
with $\nabla_\mu g_{\lambda\sigma} = 0$.  Furthermore, the source term
is given by
\begin{equation}
S_T = \int d^4 x \sqrt{g} \, h_{\mu\nu} T^{\mu\nu} \, .
\end{equation}
As in Ref.~\cite{Porrati}, we apply the simplifying assumption that
$T_{\mu\nu}$ is conserved with respect to the background metric,
$\nabla_{\mu} T^{\mu\nu} = 0$.

$S_L$ and $S_T$ together correspond to the linearized massless theory
coupled to a conserved source.  Each term independently has a gauge symmetry
described by a vector $\xi_\mu(x)$:
\begin{equation}
\label{lineargauge}
h_{\mu\nu} \rightarrow h_{\mu\nu}  + 2
\nabla_{(\mu} \xi_{\nu )}\, ,
\end{equation}
corresponding to diffeomorphism invariance of the Einstein theory.
Introduction of the Pauli-Fierz spin-2 mass term,
\begin{equation}
S_M = \tfrac{M^2}{2} \int \d^4x \sqrt{g}
\left[ h^{\mu\nu} h_{\mu\nu} - (h^\mu{}_\mu)^2 \right]\ ,
\end{equation}
breaks the symmetry (\ref{lineargauge}).  However, at the critical value
$M^2=2\Lambda/3$, there remains a residual symmetry
\begin{equation}
\label{desergauge}
h_{\mu\nu} \rightarrow h_{\mu\nu}
+ 2 \nabla_{(\mu} \nabla_{\nu )} \alpha
+ \tfrac{2}{3}\Lambda g_{\mu\nu}\alpha
\end{equation}
parameterized by $\alpha(x)$.  This gauge invariance was first noted in
\cite{deser4}, and results in a partially massless de Sitter theory with
four degrees of freedom and propagation along the light cone.  It 
also requires that the coupling to matter be via a tracelees 
energy-momentum tensor.

We wish to consider the generating functional
\begin{equation}
\label{Z}
Z[g,T] = \int \D h \, e^{-\left( S_L[h] + S_M[h] + S_T[h\cdot T]\right)}\, .
\end{equation}
Since the generic theory with mass term has broken gauge invariance and
a quadratic action, it may be quantized in a straightforward manner.
On the other hand, for the case $M^2=2\Lambda/3$, one would first gauge
fix the symmetry (\ref{desergauge}) before proceeding.  However, to make
contact with previous results for the pure massless case, we find it
useful to reintroduce the gauge symmetry (\ref{lineargauge}) using a
St\"uckelberg \cite{stuck,Higuchi} formulation.  This allows a uniform
approach to quantization throughout the $(\Lambda,M^2)$ plane, and provides
connection to the operators appearing in Ref.~\cite{Christensen} for the
massless case, as well as the ones in Ref.~\cite{dilkes} for the massive
case.


For any value of $M^2>0$, we introduce an auxiliary vector field $V_\mu$
to restore the gauge symmetry (\ref{lineargauge}).
We first multiply $Z[g,T]$ by an integration $\int \D V$ over all
configurations of this decoupled field, and then perform the shift
$h_{\mu\nu} \rightarrow h_{\mu\nu} - 2M^{-1} \nabla_{(\mu}  V_{\nu )}$.
Since $S_L$ and $S_T$ are gauge invariant in themselves, the only effect
of this shift is to make the replacement
\begin{equation}
S_M[h_{\mu\nu}]\to S_M[h_{\mu\nu} - 2 M^{-1} \nabla_{(\mu} V_{\nu )}]
\end{equation}
in (\ref{Z}).  Thus $S_M$ becomes a ``St\"uckelberg mass'', and gauge
invariance is restored, yielding the simultaneous shift symmetry
\begin{eqnarray}
\label{lineargauge2}
h_{\mu\nu}&\rightarrow&h_{\mu\nu} + 2 \nabla_{(\mu} \xi_{\nu )}\, ,\nonumber\\
V_\mu&\rightarrow&V_\mu + M \xi_\mu \, .
\end{eqnarray}

For generic $M^2$, this is the only symmetry of theory.  However,
for $M^2=2\Lambda/3$, the additional symmetry (\ref{desergauge}) remains
even after the St\"uckelberg shift.  Note that this symmetry is a
combination of a Weyl scaling and diffeomorphism [with parameter
$\xi_\mu(x)=\nabla_\mu\alpha(x)$].  Since the latter has been restored
by the addition of $V_\mu$, we are now able to disentangle the two.
The resulting gauge symmetry for the partially massless theory may be
written as
\begin{eqnarray}
h_{\mu\nu}&\rightarrow&h_{\mu\nu} + 2\nabla_{(\mu} \xi_{\nu )(x)}
+ \tfrac{2}{3}\Lambda g_{\mu\nu}\alpha(x)\, ,\nonumber\\
V_{\mu}&\rightarrow&V_\mu + M[\xi_{\mu}(x) - \nabla_{\mu} \alpha(x)]
\label{eq:totgauge}
\end{eqnarray}
with parameters $\xi_{\mu}(x)$ for diffeomorphisms and $\alpha(x)$ for
Weyl rescalings.


For the partially massless theory, there are five degrees of freedom to
gauge fix.  As in \cite{dilkes}, we make use of diffeomorphisms to
identify $V$ with the longitudinal part of $\tilde h$,
{\it i.e.}\ $M V_\mu = \nabla^\rho \tilde h_{\rho \mu}$. Additionally,
the conformal rescaling may be used to make $h_{\mu\nu}$ traceless. This
choice is made in order to simplify the relevant operators appearing in
the action, and is accomplished by adding to the action the gauge-fixing
terms
\begin{eqnarray}
S_{\rm gf}&=&\int \d^4 x \sqrt{g}
\left( \nabla^\rho \tilde h_{\rho\mu} - M V_{\mu} \right)
\left( \nabla^\sigma \tilde h_\sigma^\mu - M V^{\mu} \right)\nonumber\\
&&+ \tfrac{2}{3} \Lambda \int \d^4 x \sqrt{g} h^2 \, .
\end{eqnarray}

In conjunction with this gauge fixing, it is necessary to include a
Faddeev-Popov determinant connected with the variation of the gauge
condition under (\ref{eq:totgauge}).
It is straightforward to show that the appropriate determinant is
\begin{equation}
\label{vectdet}
\Det\pmatrix{
\left[ \Delta\halfhalf - 4\Lambda/3 \right]\  & 0 \cr
2 \nabla_{\mu} & 8\Lambda/3\cr
}
\end{equation}
corresponding to the set of gauge parameters $(\xi_{\mu}, \alpha)$.
So up to an overall (infinite) constant piece, $\Det[8\Lambda/3]$, the
relevant Faddeev-Popov term is $\Det[ \Delta\halfhalf - 4\Lambda/3 ]$
where the second-order vector spin operator is defined by
$\Delta\halfhalf \xi_\mu \equiv - \Box \xi_\mu + R_{\mu\nu} \xi^\nu$
\cite{Christensen}, and we have exploited the Einstein condition
for the background metric.  Note that, after gauge fixing, there remains
a coupling proportional to $h^\sigma{}_\sigma \nabla \cdot V$ which can
be eliminated by making the change of variables $V_\mu \rightarrow V_\mu +
\left({M\over-4 \Lambda + 2M^2}\right) \nabla_\mu h^\sigma{}_\sigma$.  


To highlight the tensor structure of the gauge-fixed action, we
decompose the metric fluctuation $h_{\mu\nu}$ into its traceless and
scalar parts: $\phi_{\mu\nu} \equiv h_{\mu\nu} - \frac{1}{4} g_{\mu\nu}
h^\sigma{}_\sigma$, and $\phi \equiv h^\sigma{}_\sigma$.  The source may
similarly be split into its irreducible components $j_{\mu\nu}$ and $j$,
so that $T_{\mu\nu} = j_{\mu\nu} + \frac{1}{4} g_{\mu\nu} j $.
The gauge-fixed partially massless action then becomes
\begin{eqnarray}
\label{Sint1}
\tilde S
& = & \int \d^4 x \sqrt{g} \Bigl[
\tfrac{1}{2} \phi^{\mu\nu} \Bigl( \Delta(1,1)
- 4\Lambda/3 \Bigr) \phi_{\mu\nu} \nonumber \\
& & \hspace{1.6cm}
+ V^\mu \Bigl(\Delta\halfhalf -4 \Lambda/3 \Bigr) V_\mu
- (\nabla \cdot V)^2
\nonumber \\
& & \hspace{1.6cm}
+ \phi_{\mu\nu} j^{\mu\nu} + \tfrac{1}{4} \phi j \Bigr] \, .
\end{eqnarray}
The second-order spin operators are the scalar Laplacian
$\Delta(0,0) \equiv - \Box$ and the Lichnerowicz operator
for symmetric rank-2 tensors
$\Delta(1,1) \phi_{\mu\nu} = - \Box \phi_{\mu\nu} + R_{\mu\tau}\phi^\tau_\nu
+ R_{\nu\tau} \phi_\mu^\tau - 2 R_{\mu\rho\nu\tau}\phi^{\rho\tau}$
\cite{Christensen}.  


The St\"uckelberg field, $V_\mu$, in (\ref{Sint1}) appears as a massive
spin-1 field in the Einstein background with an effective mass
$m^2 = -4\Lambda/3$.  We now restore vector gauge invariance by
repeating the St\"uckelberg formalism.  Thus we introduce a scalar
field $\chi$ and make the change of variables
$V_\mu \rightarrow V_\mu - M^{-1} \nabla_\mu \chi $.
By construction, the resulting action is now invariant  under the
gauge transformation
\begin{eqnarray}
V_\mu & \rightarrow & V_\mu + \nabla_\mu \zeta\, , \nonumber\\
\chi & \rightarrow & \chi + M \zeta \, .
\end{eqnarray}
One can then choose a gauge-condition to simplify the shifted action.
It is useful to associate the longitudinal component of $V$ with $\chi$
according to $M\nabla \cdot V = (-2 \Lambda + M^2) \chi$.  This is done
by adding a gauge-fixing term
\begin{equation}
\label{eq:gf2}
S_{\rm gf}' = \int \d^4 x \sqrt{g}
\left(\nabla \cdot V - \tfrac{-2 \Lambda + M^2}{M} \chi \right)^2 \, ,
\end{equation}
along with a corresponding scalar Faddeev-Popov determinant
\begin{equation}
\label{FPprimed}
\Det\left[\Delta(0,0) - 2 \Lambda + M^2\right] \, .
\end{equation}

The final completely gauged-fixed action for the partially massless graviton
now takes the form
\begin{eqnarray}
\label{Sintf}
\tilde S
& = & \int \d^4 x \sqrt{g} \Bigl[
\tfrac{1}{2} \phi^{\mu\nu} \Bigl( \Delta(1,1)
- 4\Lambda/3 \Bigr) \phi_{\mu\nu} \nonumber \\
& & \hspace{1.6cm}
+ V^\mu \Bigl(\Delta\halfhalf -4\Lambda/3 \Bigr) V_\mu \nonumber \\
& & \hspace{1.6cm}
-2 \chi \Bigl(\Delta(0,0) - 4\Lambda/3 \Bigr) \chi \nonumber \\
& & \hspace{1.6cm}
+ \phi_{\mu\nu} j^{\mu\nu} + \tfrac{1}{4} \phi j \Bigr] \, .
\end{eqnarray}
Along with the addition to the two Faddeev-Popov determinants
(\ref{vectdet}) and (\ref{FPprimed}), this provides a complete
description of $Z$, including couplings to the background metric.
This is to be compared with the generic massive 
case where the corresponding action is given by\cite{dilkes}
\begin{eqnarray}
\label{Sint2}
\tilde S
& = & \int \d^4 x \sqrt{g} \Bigl[
\tfrac{1}{2} \phi^{\mu\nu} \Bigl( \Delta(1,1)
- 2 \Lambda + M^2 \Bigr) \phi_{\mu\nu} \nonumber \\
& & \hspace{1.6cm}
- \tfrac{1}{8} \left( \tfrac{-2 \Lambda + 3 M^2}{-2 \Lambda + M^2} \right)
 \phi \Bigl( \Delta(0,0) -
2 \Lambda + M^2 \Bigr) \phi \nonumber\\
& & \hspace{1.6cm}
+ V^\mu \Bigl(\Delta\halfhalf -2 \Lambda + M^2\Bigr) V_\mu \nonumber\\
& & \hspace{1.6cm}
+\left(\tfrac{-2 \Lambda + M^2 }{M^2}\right)
\chi \Bigl(\Delta(0,0) - 2 \Lambda + M^2 \Bigr) \chi \nonumber \\
& & \hspace{1.6cm}
+ \phi_{\mu\nu} j^{\mu\nu} + \tfrac{1}{4} \phi j \Bigr] \, .
\end{eqnarray}

Note that in the partially massless case the trace mode $\phi$ has 
disappeared except for its coupling to the trace of the energy-momentum 
tensor.  With $\phi$ now acting as a Lagrange multiplier, this indicates 
that the theory couples to conformal matter.  To compare the massive 
and partially massless theories at the classical level, therefore, let 
us assume that the massive theory also couples to matter with 
$T^\mu_\mu=0$ as well as $\nabla_{\mu} T^{\mu\nu} = 0$. Then 
the tree-level amplitude for the current $T_{\mu\nu}$
can be read from the action (\ref{Sint2}) directly and is given by
\begin{eqnarray}
\label{amp} A[T] & = & \tfrac{1}{2} \, T^{\mu\nu} 
\left(\Delta(1,1)-2\Lambda+M^2 \right)^{-1} T_{\mu\nu} \, ,\nonumber
\end{eqnarray}
since there are sources for neither $V_{\mu}$ nor $\chi$.  Thus at 
tree level, there is no discontinuity in taking the $M^{2} \rightarrow 
2\Lambda/3$ limit.  We note here that there would be sources for the
St\"uckelberg fields
if one were to relax the assumption of a conserved stress tensor or a 
traceless stress tensor.  In this case, one needs only to account for the 
shifts in $h_{\mu\nu}$ and $V_\mu$ to see how $T_{\mu\nu}$ contributes 
to sources for $V_\mu$ and $\chi$.

For the partially massless case, (\ref{Sintf}), we integrate over all
species to find the first quantum correction 
\ifpreprintsty\else
\end{multicols}
\narrowtext
\hbox to \hsize{\hrulefill}
\fi
\widetext
\begin{eqnarray}
Z[g,T]  \propto
e^{-A[T]}
&&\Det \Bigl[\Delta\halfhalf - \tfrac{4}{3}\Lambda \Bigl]
\;\Det \Bigl[\Delta(0,0) - \tfrac{4}{3}\Lambda \Bigl] \nonumber\\   
\times&&
\Det \Bigl[\Delta(1,1) - \tfrac{4}{3}\Lambda \Bigl]^{-1/2}
\Det \Bigl[\Delta\halfhalf - \tfrac{4}{3}\Lambda  \Bigl]^{-1/2}
\Det \Bigl[\Delta(0,0) - \tfrac{4}{3}\Lambda \Bigl]^{-1/2} \kern-1.5em ,
\end{eqnarray}
where the operator $\Delta(1,1) -\tfrac{4}{3}\Lambda$ arises in the
traceless $\phi^{\mu\nu}$ sector so its determinant refers
to traceless modes only.
This allows us to compute the one-loop contribution
\begin{equation}
\Gamma^{(1)}[g] = - \ln Z[g,0]
= - \tfrac{1}{2} \ln \Det \Bigl[\Delta\halfhalf - \tfrac{4}{3} \Lambda
\Bigl]
+ \tfrac{1}{2} \ln \Det \Bigl[\Delta(1,1) - \tfrac{4}{3} \Lambda  \Bigl]
- \tfrac{1}{2} \ln \Det \Bigl[\Delta(0,0) - \tfrac{4}{3} \Lambda \Bigl]   
\ifpreprintsty\kern-23pt\fi
\end{equation}
to the effective action for the Einstein background $g_{\mu\nu}$. This is
now to be compared with the one loop contribution in the generic massive
case \cite{dilkes}
\begin{equation}
\label{genericgamma}
\Gamma^{(1)}[g] = - \ln Z[g,0] = -\tfrac{1}{2} \ln \Det 
\Bigl[\Delta\halfhalf - 2 \Lambda + M^2 \Bigl] + \tfrac{1}{2} \ln \Det 
\Bigl[\Delta(1,1) - 2 \Lambda + M^2 \Bigl].
\end{equation}

The difference in these two expressions reflects the fact that 5
degrees of freedom are being propagated around the loop in the massive
case and only 4 in the partially massless case. Denoting the dimension of
the spin $(A,B)$ representation by $D(A,B)=(2A+1)(2B+1)$,
we count $D(1,1)-D(1/2,1/2)=5$ for the massive case, while
$D(1,1)-D(1/2,1/2)-D(0,0)=4$ for the partially massless one.

It remains to check that there is no conspiracy among the eigenvalues of
these operators that would make these two expressions coincide. To show
this, it suffices to calculate the coefficients in the heat-kernel
expansion for the graviton propagator associated with $S_L + S_M$,
and compare it with the massive case given in Ref.~\cite{dilkes}.
The coefficient functions $b_k^{(\Lambda)}$ in the expansion
\begin{equation}     
\Tr e^{-\Delta^{(\Lambda)} t}
= \sum_{k=0}^\infty t^{(k-4)/2}
\int \d^4x \sqrt{g} \, b_k^{(\Lambda)}
\end{equation}
were calculated in Ref.~\cite{Christensen} for general ``spin operators''
$\Delta^{(\Lambda)}(A,B)\equiv\Delta(A,B) -2\Lambda$ in an Einstein 
background $R_{\mu\nu}=\Lambda g_{\mu\nu}$.  So, adapted for the generic 
operators $\Delta^({\Lambda,M})\equiv \Delta(A,B) -2\Lambda+M^{2}$ 
appearing in Eq.~(\ref{genericgamma}), but still in the same Einstein 
background $R_{\mu\nu}=\Lambda g_{\mu\nu}$, the results are:
\begin{eqnarray}
180(4\pi)^2 b_4^{(\Lambda,M)}(1,1)&=&189R_{\mu\nu\rho\sigma} 
R^{\mu\nu\rho\sigma}-756\Lambda^{2}+810M^{4}\, ,\nonumber\\
180(4\pi)^2 b_4^{(\Lambda,M)}(\half,\half)&=&-11R_{\mu\nu\rho\sigma} 
R^{\mu\nu\rho\sigma}+984\Lambda^{2}-1200\Lambda M^{2}+360M^{4}\, 
,\nonumber\\
180(4\pi)^2 b_4^{(\Lambda,M)}(0,0)&=&R_{\mu\nu\rho\sigma} 
R^{\mu\nu\rho\sigma}+636\Lambda^{2}-480\Lambda M^{2}+90M^{4}\, .
\end{eqnarray}
\ifpreprintsty\else
\narrowtext \dimen0\hsize
\widetext
\hbox to \hsize{\hss \hbox to \dimen0{\hrulefill}}
\begin{multicols}{2}
\fi
\narrowtext

For the partially massless four degrees of freedom theory,
$M^2=2\Lambda/3$, we obtain
\begin{eqnarray}
180(&&4\pi)^2b_4^{(\Lambda)}({\rm total})\nonumber\\
&&= 180 (4\pi)^2 \left[b_4^{(\Lambda)}(1,1)-b_4^{(\Lambda)}(\half,\half)
-b_4^{(\Lambda)}(0,0)\right]\nonumber\\ 
&&= 199 R_{\mu\nu\rho\sigma} R^{\mu\nu\rho\sigma} - 1096 \Lambda^2 \, ,
\end{eqnarray}
which differs from the result for the $M^{2}\rightarrow 2 \Lambda/3$
limit of the massive case,
\begin{eqnarray}
180(4\pi)^2b_4^{(\Lambda,M)}&&({\rm total})\nonumber\\
&&= 180(4\pi)^2 \left[
b_4^{(\Lambda,M)} (1,1) - b_4^{(\Lambda,M)}\halfhalf \right]\nonumber\\
&& \rightarrow
200 R_{\mu\nu\rho\sigma} R^{\mu\nu\rho\sigma} -740 \Lambda^2 \, .
\end{eqnarray}
Even for a de Sitter background with constant curvature
\begin{eqnarray}
R_{\mu\nu\rho\sigma}&=&
\tfrac{1}{3}\Lambda(g_{\mu\nu}g_{\rho\sigma}-g_{\mu\rho}g_{\nu\sigma})
\, ,\nonumber\\
R_{\mu\nu\rho\sigma}R^{\mu\nu\rho\sigma}&=&\tfrac{8}{3}\Lambda^{2}\, ,
\end{eqnarray}
there is no cancellation.

Thus we conclude that the absence of a discontinuity between the
$M^2 \rightarrow 2\Lambda/3$ and $M^2=2\Lambda/3$ results for massive
spin-2 is only a tree-level phenomenon, and that the discontinuity
itself persists at one loop.
That the full quantum theory is discontinuous is not surprising
considering the different degrees of freedom for the two cases.  Just 
as the $ M^{2}\rightarrow 0$ limit is discontinuous at the quantum level 
as a result of going from five degrees of freedom to two, so the $M^2 
\rightarrow 2\Lambda/3$ limit is discontinuous as a result of going 
from five degrees of freedom to four.

\section*{acknowledgments}

We wish to thank S.~Deser for enlightening discussions and F.~Dilkes for
initial collaboration. This research was 
supported in part by DOE Grant DE-FG02-95ER40899.


\ifpreprintsty\else
\end{multicols}
\fi


\begin{references}

\bibitem{dilkes}
F.~A.~Dilkes, M.~J.~Duff, James T.~Liu and H.~Sati,
\rtitle{Quantum $M \rightarrow 0$ discontinuity for massive gravity
with a $\Lambda$ term}.
 hep-th/0102093.

\bibitem{Higuchi}
A.~Higuchi,
\rtitle{Forbidden Mass Range For Spin-2 Field Theory In De Sitter
Space-Time}
Nucl.\ Phys.\ {\bf B282}, 397 (1987).

\bibitem{Kogan}
I.~I.~Kogan, S.~Mouslopoulos and A.~Papazoglou,
\rtitle{The $m \to 0$ limit for massive graviton in $dS_4$ and $AdS_4$:
        How to circumvent the van Dam-Veltman-Zakharov discontinuity}
Phys.\ Lett.\ B {\bf 503}, 173 (2001)
[hep-th/0011138].

\bibitem{Porrati}
M.~Porrati,
\rtitle{No van Dam-Veltman-Zakharov discontinuity in AdS space}
Phys.\ Lett.\ B {\bf 498}, 92 (2001)
[hep-th/0011152].

\bibitem{vdv}
H.~van Dam and M.~Veltman,
\rtitle{Massive and massless Yang-Mills and gravitational fields}
Nucl.\ Phys.\ {\bf B22}, 397 (1970).

\bibitem{zak}
V.I. Zakharov, JETP Lett. {\bf 12}, 312 (1970).

\bibitem{Deser:2001pe}
S.~Deser and A.~Waldron,
\rtitle{Gauge invariances and phases of massive higher spins in (A)dS}
hep-th/0102166.

\bibitem{deser4}
S.~Deser and R.~I.~Nepomechie,
\rtitle{Gauge Invariance Versus Masslessness In De Sitter Space}
Annals Phys.\ {\bf 154}, 396 (1984).

\bibitem{deser}
S.~Deser and A.~Waldron,
\rtitle{Partial Masslessness of Higher Spins in (A)dS}
hep-th/0103198.

\bibitem{Deser:2001wx}
S.~Deser and A.~Waldron,
\rtitle{Stability of massive cosmological gravitons}
hep-th/0103255.

\bibitem{stuck}
E.C.G. St\"uckelberg, Helv. Phys. Acta {\bf 30}, 209 (1957).

\bibitem{Christensen}
S.~M.~Christensen and M.~J.~Duff,
\rtitle{Quantizing gravity with a cosmological constant}
Nucl.\ Phys.\ {\bf B170}, 480 (1980).

\end{references}
\end{document}